\documentstyle [epsfig, 12pt]{article}
\begin{document}
\bigskip

Romanian Astronomical Journal {\it Vol. 15, No. 1, p. 3-7,
Bucharest, 2005}
\\ \\

\bigskip

\begin{center}
{\Large {\bf Real  Size  And  Membership  Richness Determination
Of  High-Latitude  Open  Clusters}}
\end{center}

\bigskip
\begin{center}
{\bf ASHRAF LATIF TADROSS}
\bigskip
\\ {\it National Research Institute of
Astronomy and Geophysics, \\ 11421 - Helwan, Cairo, Egypt
\\E-mail: altadross@nriag.sci.eg}
\end{center}
\bigskip
{\it Abstract.} We use proper motion measurements to determine the
real size and membership richness of a sample of open clusters
located at high galactic latitudes $(40 \leq |b| \leq 90)$.

\bigskip
\begin{center}
{\bf 1. INTRODUCTION}
\end{center}

USNO-B1.0 of Monet et al. (2003) is a spatially unlimited catalog
that presents positions, proper motions, and magnitudes in various
optical pass-bands. The data were obtained from scans of Schmidt
plates taken for the various sky surveys during the last 50 years.
USNO-B is believed to provide all-sky coverage, completeness down
to V = 21 mag. It is noted that, based on the proper motion
measurements, stars with large proper motions are likely to be
foreground stars instead of cluster members. Background stars
cannot readily be distinguished from members by proper motions.
Nonetheless, identifying foreground stars is useful in cleaning up
the color-magnitude diagrams and estimating the amount of field
star contamination. So, USNO-B is a very useful catalog, which
gives us an opportunity to distinguish between the members and
field and background stars. On the other hand, the use of
proper-motion-based membership for a very distant cluster cannot
be regarded as a good idea. However, in particular cases it can
work, for clusters located at high latitude above the galactic
plane, hence contaminated by relatively nearby stars only, where
far background and foreground stars become rare. This point
however requires validation and detailed discussion due to the
peculiarities of the USNO-B catalog, which is the main idea of the
present work.

On this respect, open clusters located at high galactic latitudes
$(40 \leq |b| \leq 90)$ have been nominated from WEBDA site
\footnote{\it http://www.univie.ac.at/webda} (16 clusters; 10
below and 6 above the galactic plane). Those clusters are
contaminated by relatively nearby stars.

Because USNO-B catalog demonstrates a pronounced excess of
zero-proper-motion objects, it is found suitable enough for the
aim of this paper. So, after removing the contaminated nearby
stars, we can determine the genuine size of each cluster of our
sample regardless on the available photometry. Applying the
completeness correction for each cluster, the membership
determination can be counted.

\bigskip
\begin{center}
{\bf 2. ON THE PRESENT WORK}
\end{center}

Using USNO-B1.0 catalog of Monet et al. (2003), we can determine
the genuine borders of the clusters under investigation. For this
purpose, the data have been obtained at a preliminary radius of
about 20 arcmin from the cluster center. For all the clusters of
our sample, all stars with nonzero proper motions and those
distributed over the field with no concentration around the
cluster center have been removed.

Within concentric shells in equal incremental steps of 0.5 arcmin
from the cluster center, stellar density is performed out to the
preliminary radius. The real cluster radius is defined at the
point that covers the cluster area and reaches enough stability in
the density of the background, i.e. the difference between the
observed density profile and the background one is almost equal to
zero (see Tadross 2005). At that pointed radius, the data
incompleteness is evaluated and removed down to a specific B2
magnitude (the second blue magnitude of USNO-B catalog) with
completeness level more than $60\%$. At this value the cluster
data can be cut-off and then the membership can be counted (Sanner
et al. 1999). An example, for estimating the true radius and the
completeness limit of the membership for NGC 1498 is shown in
Figs. 1 and 2.

\begin{figure}
\begin{center}
\epsfig{file=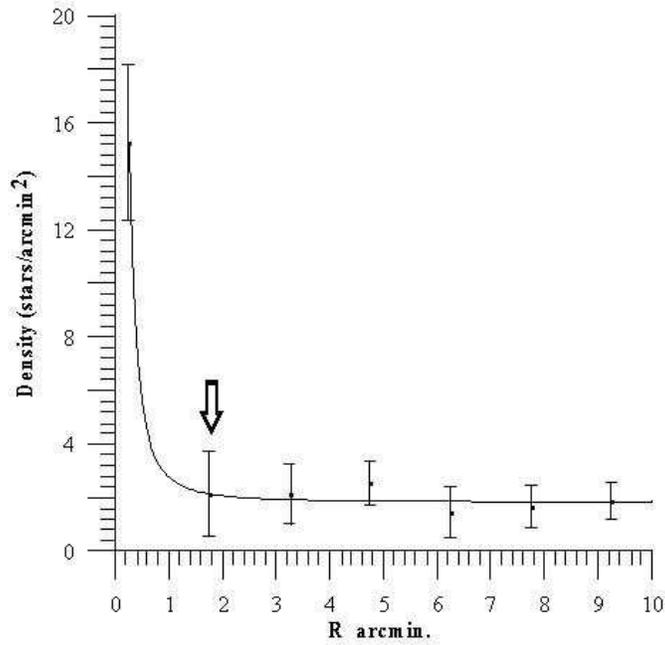,width=9 cm} \caption{\small An example for
estimating the true radius of NGC 1498 using the projected density
distribution. The lengths of the error-bars denote errors
resulting from sampling statistics, in accordance with Poisson
distribution (1/N)$^{1/2}$, where N is the number of stars used in
the density estimation at that point. The arrow represents the
point at which the radius of the cluster is obtained.}
\end{center}
\end{figure}

\begin{figure}
\begin{center}
\epsfig{file=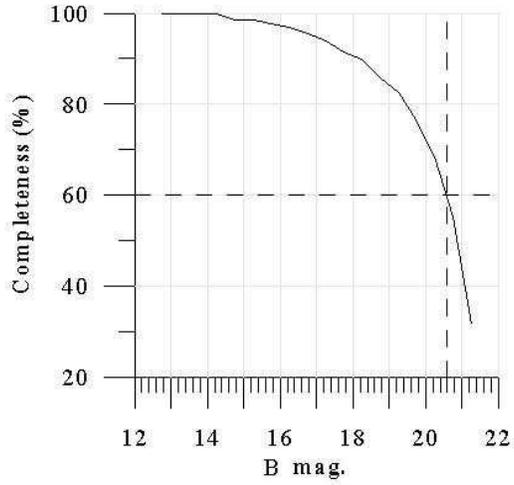,width=7 cm} \caption{\small An example for
estimating the completeness limit at which the membership of NGC
1498 has been counted. Horizontal and vertical dashed lines
represent this limit $\geq 60\%$ at 20.60 mag.}
\end{center}
\end{figure}

\begin{figure}
\begin{center}
\epsfig{file=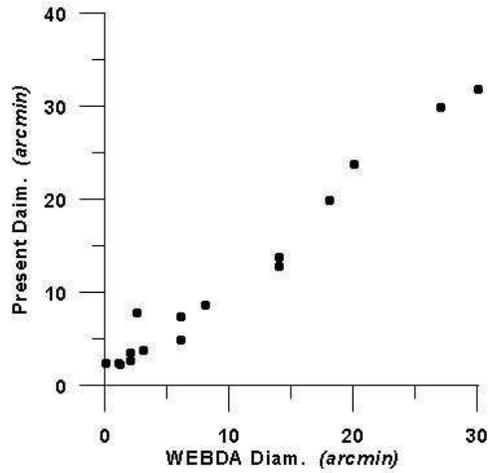,width=7 cm} \caption{\small Comparison
between the resultant diameters and those given in WEBDA. The
correlation coefficient is found to be 0.98.}
\end{center}
\end{figure}

\begin{center}
{\bf 3. CONCLUSIONS}
\end{center}
Following the above procedure, the real size, completeness limit,
and membership richness of all the clusters of our sample have
been determined and listed in Table 1. It is noted that most of
the clusters in the present work have larger diameters and contain
more members than those obtained in the literature. For the sake
of comparison, the relation between the resultant diameters and
those given in WEBDA is supposed to be linear with correlation
coefficient of 0.98, as shown in Fig. 3. On the other hand, the
resultant diameters are found related to the richness in
proportional way as shown in Fig. 4. This point was declared
before in previous studies (see Tadross et al. 2002, and the
references therein).
\\
\\
\small {\it Acknowledgment.} I would like to express my
appreciation to the teamwork of USNO-B catalog (Monet et al. 2003)
for providing their useful catalog that serves such a kind of
work.

\begin{figure}
\begin{center}
\epsfig{file=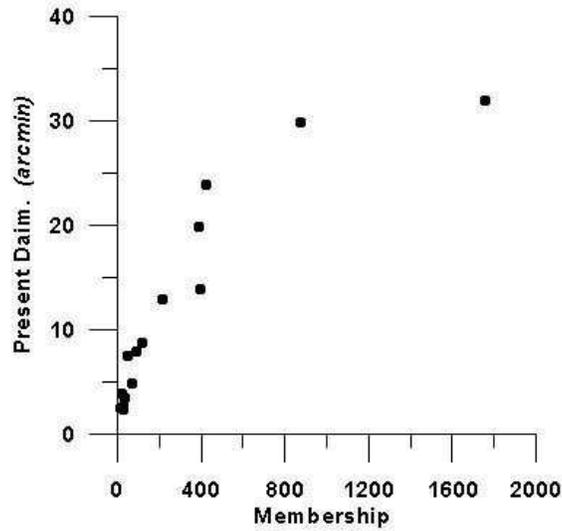,width=8 cm} \caption{\small The relation
between the resultant diameters and membership richness. It is
noted that the number increases with diameter in proportional way;
see the text.}
\end{center}
\end{figure}

\newpage
\begin{table}
\small
\begin{center}
\caption{The main parameters of 16 clusters of our sample with
completeness limit, genuine diameters and membership richness.}
\label{t1}
\begin{tabular}{llllllllll}
\hline\hline
Cluster & $\alpha_{~2000}$ & $\delta_{~2000}$ & l $_{2000}$ & b $_{2000}$ &  C.L.  & Diam. & Mem.\\
        & h \, m \, s  & $^{\circ} \, \, \, \, \, \, \, \, ' \, \, \, \, \, \, \, \, ''$  & {\it deg}  & {\it deg}  &  & {\it arcmin} \\
\hline
Upgren 1 & 12 35 00 & +36 18 00 & 142.740 & 80.188 & 20.5 & 20 & 382 \\
Chereul 1 & 14 29 04 & +55 23 30 & 97.639 & 56.679 & 20.9 & 2.6 & 23 \\
NGC 3231 & 10 26 58 & +66 48 42 & 141.924 & 44.653 & 20.6 & 8 & 82 \\
Dol-Dzim 5 & 16 27 24 & +38 04 00 & 60.866 & 43.875 & 20.5 & 30 & 870 \\
NGC 5385 & 13 52 27 & +76 10 24 & 118.197 & 40.389 & 21 & 8.8 & 115 \\
Dol-Dzim 6 & 16 45 24 & +38 21 00 & 61.583 & 40.368 & 20.3 & 5 & 62 \\
AM 0430-392 & 04 32 24 & -39 17 36 & 242.616 & -42.958 & 20.9 & 2.8 & 25 \\
NGC 1498 & 04 00 18 & -12 00 54 & 203.623 & -43.330 & 20.6 & 3.6 & 27 \\
ESO 236-07 & 21 21 28 & -51 48 42 & 345.827 & -43.901 & 20.5 & 32 & 1750 \\
NGC 7772 & 23 51 46 & +16 14 48 & 102.740 & -44.273 & 20.8 & 4 & 18 \\
NGC 7134 & 21 48 55 & -12 58 24 & 41.980 & -45.141 & 20.4 & 2.6 & 11 \\
NGC 305 & 00 56 20 & +12 04 00 & 124.825 & -50.787 & 20.6 & 7.6 & 40 \\
NGC 1252  & 03 10 49 & -57 46 00 & 274.084 & -50.831 & 20.7 & 14 & 390 \\
Whiting 1 & 02 02 57 & -03 15 10 & 161.618 & -60.636 & 20.2 & 2.4 & 25 \\
ESO 245-09 & 01 53 43 & -45 57 12 & 273.762 & -67.488 & 21 & 13 & 210 \\
NGC 7826 & 00 05 17 & -20 41 30 & 61.875 & -77.653 & 20.6 & 24 & 415 \\

\hline\hline
\end{tabular}
\end{center}
\end{table}

\begin{center}
{\bf REFERENCES}
\end{center}
\bigskip
Monet, D. et al.: 2003, {\it Astron. J.}, {\bf 125}, 984.
\\
Sanner, J., Geffert, M., Brunzendorf, J., Schmoll, J.: 1999, {\it
Astron. Astrophys.}, {\bf 349}, 448.
\\
Tadross, A. L.: 2005, {\it Astron. Nachr.}, {\bf 326}, 19.
\\
Tadross, A. L., Werner, P., Osman, A., Marie, M.: 2002, {\it New
Astron.}, {\bf 7}, 553.

\end{document}